\documentclass[aps,prl,showpacs,twocolumn,showkeys,amssymb,amsmath,nobibnotes,nofootinbib,superscriptaddress,reprint,longbibliography]{revtex4-1}
\usepackage{verbatim}
\usepackage{amssymb}
\usepackage{graphicx}
\usepackage{subfigure}
\usepackage[colorlinks=true,linkcolor=red]{hyperref}
\usepackage[sort&compress]{natbib}
\usepackage{color}

\begin{document}

\title{Interacting ultracold atomic kicked rotors: dynamical localization ?}

\author{Pinquan Qin}
\author{Alexei Andreanov}
\author{Hee Chul Park}
\affiliation{Center for Theoretical Physics of Complex Systems, Institute for Basic Science, Daejeon, South Korea}

\author{Sergej Flach}
\affiliation{Center for Theoretical Physics of Complex Systems, Institute for Basic Science, Daejeon, South Korea}
\affiliation{New Zealand Institute for Advanced Study, Center for Theoretical Chemistry \& Physics, Massey University, Auckland, New Zealand}

\begin{abstract}
    We study the fate of dynamical localization of two quantum kicked rotors with contact interaction. This interaction mimics experimental realizations with ultracold atomic gases. Dynamical localization for a single rotor takes place in momentum space. The contact interaction affects the evolution of the relative momentum $k$ of a pair of interacting rotors in a non-analytic way. Consequently the evolution operator $U$ is exciting large relative momenta with amplitudes which decay only as a power law $1/k^4$. This is in contrast to the center-of-mass momentum $K$ for which the amplitudes excited by $U$ decay superexponentially fast. Therefore dynamical localization is preserved for the center-of-mass momentum, but destroyed for the relative momentum for any nonzero strength of interaction.
\end{abstract}

\maketitle

The quantum kicked rotor (QKR) model is a canonical model to explore quantum chaos~\cite{chirikov1979universal,izrailev1990simple}. It describes a quantum rotor degree of freedom which is periodically kicked by a force periodic in the angle. The QKR enjoys dynamical localization (DL) - i.e. the arresting of the growth of the momentum despite the absence of a cutoff in the frequency of the kick drive. DL was first discovered numerically by Casati, Chirikov, Ford, and  Izrailev~\cite{casati1979stochastic} and later confirmed experimentally for Rydberg atoms in a microwave  field~\cite{galvez1988microwave,bayfield1989localization} and ultracold atomic gases in a modulated standing wave of a near-resonant laser~\cite{moore1994observation}. A recent work reports on the experimental observation of DL with laser-kicked molecular rotors~\cite{bitter2016experimental}. If the driving period is an irrational multiple of $2\pi$, the rotor is localized in the momentum space, even though the classical counterpart shows diffusive momentum growth. This happens because classical chaotic diffusion is suppressed by quantum interference effects. The mechanism of DL was described in a seminal paper by Fishman, Grempel and Prange~\cite{fishman1982chaos}. These authors demonstrated that the kicked rotor model maps directly to an Anderson-like model with a quasi-periodic potential, which originates from the irrational driving periods. Therefore DL is closely related to Anderson localization of waves in truly random (uncorrelated) potentials.

The original quantum kicked rotor corresponds to a single quantum particle problem. The effect of interactions on Anderson localization has been attracting a lot of interest recently and several theoretical studies considered various versions of interacting kicked rotors. In~\cite{keser2015dynamical} a similar problem was studied for a simpler, integrable model of linear rotors~\cite{fishman1982chaos}, where localization can survive in the presence of interactions due to integrability. The authors of ~\cite{rozenbaum2016dynamical} analyzed coupled relativistic rotors which might be applicable to fermions in pulsed magnetic fields, and report that DL can be destroyed by suitable parameter tuning. In~\cite{nag2005quantum}, two kicked rotors with product sinusoidal interaction at the kick were studied with respect to temporal fluctuations in the reduced density matrix.  In~\cite{adachi1988quantum}, the coupling was sinusoidal depending on the two rotors relative coordinates: recovering of the chaotic behavior was found above some  kicking threshold in the semi-classical approximation. In~\cite{toloui2009quantum}, the interaction at the kick of the kicked rotors contained both product and relative coordinate dependent sinusoidal terms. Localization was found for weak coupling and quasi-diffusive regime was found for stronger interaction with a complex intermediate regime. 
 
From the experimental perspective, interaction between rotors is negligible for Rydberg atoms and laser-kicked molecular rotors. However the interaction between ultracold atoms in a Bose-Einstein condensate can be substantial, and even tunable using Feshbach resonances~\cite{chin2010feshbach}, which is particularly true for sodium atoms used in \cite{moore1994observation}. The atom-atom interaction in this case is typically of a contact type, i.e. the atoms interact through a $\delta(x_1-x_2)$ potential~\cite{chin2010feshbach}. For the experimental realization in \cite{moore1994observation} this interaction persists at all times - in contrast to the kick potential, and in contrast to the theoretical studies discussed above, which consider a kicked (time-dependent) interaction. A $\delta(x)$ interaction is long ranged in momentum space, and can therefore have a qualitatively strong impact on DL for interacting ultracold atoms. Will DL survive, or not?

In this Letter, we provide an answer to this question. We consider two bosons interacting via a $\delta$-function potential that are driven by a periodic kicking potential. The wave function for two $\delta$-function interacting bosons is computed. At variance to the Lieb-Liniger model approach~\cite{lieb1963exact}, we use the center of mass and relative coordinates with appropriate periodic boundary conditions. A repulsive $\delta$-interaction is considered, that does not lead to the appearance of a bound state. In the chosen basis the matrix elements of the time evolution operator show different decay rates along the center of mass (superexponential) and the relative momentum (algebraic) directions. Due to this qualitative difference in the decay properties of the matrix elements, dynamical localization is destroyed for the relative momentum, while being preserved for the center-of-mass momentum.

We consider two bosons moving on a ring $[0, 2\pi)$ with $\delta$-function interaction and periodic kicking potential. The Hamiltonian of the model is given by:
\begin{equation}
    \label{Ht}
    H = H_\delta + H_k\sum_n\delta\left(\frac{t}{T} - n\right),
\end{equation}
where $H_\delta = p_1^2/2M + p_2^2/2M + \lambda\delta(x_1 - x_2)$, $H_k = \xi\left[\cos(x_1) + \cos(x_2)\right]$; $\lambda$ is the interaction strength, $\xi$ is the kicking strength. This system corresponds to accounting for the atom-atom interaction in the experimental setup in~\cite{moore1994observation} for two atoms, as a first step towards the consideration of a many body interacting system. We can therefore view our model as a simple paradigmatic case of just two interacting atoms which is the building block of reaching out to many body interactions.

We start by computing the wave function of the two bosons system with $\delta$-function interaction. It can be represented  in a center of mass and relative coordinates frame - ($y_1$, $y_2$), where $y_1 = x_1 + x_2$, $y_2 = x_1 - x_2$. In this frame, the first part of~\eqref{Ht} reads
\begin{equation}
    \label{hdelta}
    H_\delta = -\frac{\hbar^2}{M}\left[\frac{\partial^2 }{\partial y_1^2} + \frac{\partial^2}{\partial  y_2^2}\right] + \lambda\delta(y_2)\;.
\end{equation}
It splits in two parts, $H_\delta = H_{y_1} + H_{y_2}$, where $H_{y_1} = -\left(\hbar^2/M\right)\left(\partial^2/\partial y_1^2\right)$, $H_{y_2} = -\left(\hbar^2/M\right)\left(\partial^2/\partial y_2^2\right) + \lambda\delta(y_2)$. $H_{y_1}$ describes a  free moving particle and $H_{y_2}$ describes a single particle with $\delta$-function potential. The wave function of the complete system is $\phi(y_1, y_2)= \phi_1(y_1) \phi_2(y_2)$ - the product of two single particle wave functions $\phi_1$ and $\phi_2$, that satisfy $H_{y_1}\phi_1 = E_{y_1}\phi_1$, $H_{y_2} \phi_2 = E_{y_2} \phi_2$. The total eigenenergy of the system is $E = E_{y_1} + E_{y_2}$. Because of the periodicity, the complete wave function satisfies $\phi(y_1,y_2) = \phi(y_1+4\pi, y_2) = \phi(y_1,y_2+4\pi) = \phi(y_1+2\pi,y_2+2\pi)$. This can be simplified into three identities: $\phi_1(y_1) = \phi_1(y_1+4\pi)$, $\phi_2(y_2) = \phi_2(y_2+4\pi)$, $\phi_1(y_1)\phi_2(y_2) = \phi_1(y_1+2\pi)\phi_2(y_2+2\pi)$, which serve as the periodic boundary conditions for $\phi_1$ and $\phi_2$.

The wavefunction for the free moving particle is:
\begin{equation}
    \phi_1(y_1) = A_1e^{iKy_1}\;.
\end{equation}
The periodic boundary conditions $\phi_1(y_1)=\phi_1(y_1+4\pi)$ select the quantized values of $K$: $e^{ i4K\pi} = 1$, giving $K = 0,\pm 1/2,\pm 1,\pm 3/2, \dots$ The normalization condition yields $A_1=1/\sqrt{4\pi}$, and the eigenenergy $E_{y_1} = \hbar^2K^2/M$.

$H_{y_2}$ describes a massive particle on a one-dimensional ring of circumference $4\pi$ and a $\delta$-function singularity at $y_2=0$. The eigenstates of this problem can be either symmetric or antisymmetric around $y_2=0$. Since rotors are bosons, the wave function should be invariant under permutation $x_1 \leftrightarrow x_2$ and only symmetric functions $\phi_2(y_2) = \phi_2(-y_2)$ are allowed. Then the derivative is an antisymmetric function $\phi'_2(y_2) = -\phi'_2(-y_2)$. The wave function is continuous at $0$: $\phi_2(+0) = \phi_2(-0)$, but its derivative has a jump: $\phi'_2(+0) - \phi'_2(-0) \equiv 2\phi'_2(+0) = \left(M \lambda/\hbar^2\right) \phi_2(0)$. From the periodic boundary condition $\phi_1(y_1)\phi_2(y_2) = \phi_1(y_1+2\pi)\phi_2(y_2+2\pi)$,  it follows that $\phi_2(y_2) = e^{i2K\pi} \phi_2(y_2+2\pi)$ and $\phi_2(y_2)=\phi_2(y_2+4\pi)$. It is worth noting that the center of mass and relative momenta do not decouple completely, due to the boundary conditions.

With these boundary conditions it is easy to compute the wave function $\phi_2$ (see Supplementary material for more details):
\begin{gather}
    \label{psiy2}
    \phi_2(y_2) = \Bigg\{
        \begin{array}{ccc}
            2B^k_K \cos [ky_2 - (k+K)\pi] ,\ \ \mathrm{if}\ \ 0 \leq y_2 < 2\pi\\
            2B^k_K \cos [ky_2 + (k+K)\pi] ,\ \ \mathrm{if}\ \  -2\pi \leq y_2  < 0 \\
        \end{array}\\
    \label{akk}
    B^k_K = \left[\sqrt{8\pi+\frac{4}{k} \sin (2k\pi)\cos(2K\pi) }\right]^{-1}\;,\\
    \label{kz}
    2(k+K)\pi = \pi - 2\arctan\left(2 kA_{\lambda}\right)\;,
\end{gather}
with $A_\lambda=\hbar^2/M\lambda$ being a dimensionless inverse interaction strength and the eigenenergy $E_{y_2} = \hbar^2 k^2/M$. We use $A_\lambda$ to measure the strength of the interaction, to which it is inversely proportional. We attach the momenta $K$ and $k$ to the full wavefunction - $\phi_K^k(y_1, y_2)$. Now the eigenstates of two bosons with $\delta$-function interaction can be written as $|\phi_K^k\rangle$, the corresponding wave function is
\begin{equation}
    \label{phikk}
    \phi_K^k(y_1,y_2) = \left\{
        \begin{array}{ccc}
            \sqrt{\frac{1}{\pi}} B^k_K\cos [ky_2 - (k+K)\pi] e^{iKy_1},\\
            \mathrm{if}\ \ 0 \leq y_2 < 2\pi\\
            \sqrt{\frac{1}{\pi}} B^k_K\cos [ky_2 + (k+K)\pi] e^{iKy_1},\\
            \mathrm{if}\ \ -2\pi \leq y_2  < 0 \\
        \end{array}\right.
\end{equation}
The eigenenergy is given by $E_K^k = \hbar^2\left(K^2 + k^2\right)/M$ These eigenstates can be used as the basis of the Hilbert space. The wavefunctions are symmetric with respect to $k$: if $k$ is a solution of Eq.~\eqref{kz} then $-k$ is also the solution. As follows from Eq.~\eqref{psiy2}, the wave function with $k$ and $-k$ are exactly the same for integer $K$ or differ by a global phase $\pi$ for half integer $K$. Consequently $\phi_K^{-k}(y_1,y_2)$ is equivalent to $\phi_K^{k}(y_1,y_2)$, reflecting the bosonic nature of the rotors. In the following discussion, we only consider the $k>0$ case.

In the presence of a periodic driving potential in Eq.~\eqref{Ht}, the dynamics is described by the time evolution operator $U$ (Floquet propagator) over one period~\cite{grifoni1998driven}. Given some initial state $|\psi(0)\rangle$ of the rotors, the final state after $N$ driving periods is
\begin{equation}
    \label{timeevo}
    |\psi(NT)\rangle = \left[U\right]^N |\psi(0)\rangle\;,
\end{equation}
For periodically kicked interacting rotors the Floquet operator reads
\begin{equation}
    U = e^{-\frac{i}{\hbar}H_{\delta}T}e^{-\frac{i}{\hbar}H_kT}\;,
\end{equation}
where $H_\delta$ is given by Eq.~\eqref{hdelta} and $H_k = 2\xi\cos(y_1/2)\cos(y_2/2)$.

\begin{figure}
	\centering
	\includegraphics[width=1.0\columnwidth]{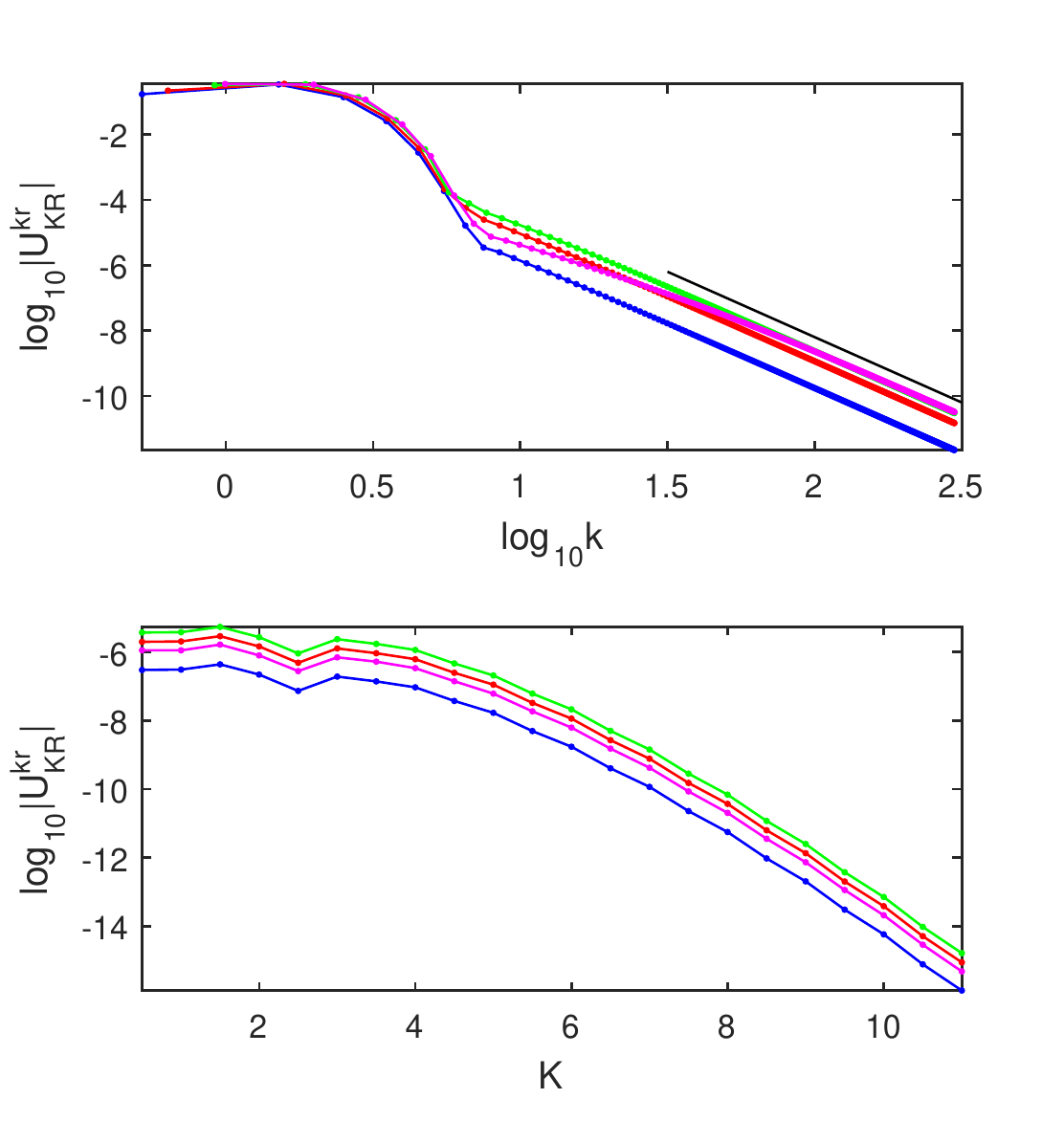}
	\caption{(Color online) Kicking strength $\xi=3$ and driving period $T=1$. The colors correspond to different inverse interactions strengths: blue - $A_\lambda=10$, red - $A_\lambda=1$, green - $A_\lambda=0.1$, magenta - $A_\lambda=0.01$. \textit{Top}: $\log_{10} |U_{KR}^{kr}|$ vs. $\log_{10} k$ with $K=R=0$ and fixed $r$ - the same values as for the bottom figure. The black line is plotted as $\log_{10} |U_{KR}^{kr}| = -4\log_{10} k-0.2$. \textit{Bottom}: $\log_{10} |U_{KR}^{kr}|$ vs. $K$ with $R=0$, fixed values of relative momenta: blue - $k=14.001$, $r=0.123$; red - $k=14.011$, $r=0.319$; green - $k=14.108$, $r=0.470$; magenta - $k=14.410$, $r=0.496$. }
	\label{matrixdecay}
\end{figure}

In the basis of $|\phi_K^k\rangle$, the matrix elements of $U$ become
\begin{gather}
    \label{ukK}
    U_{KR}^{kr} \equiv \langle \phi_R^r|U|\phi_K^k\rangle = e^{-\frac{i}{\hbar}E_R^r T}\langle \phi_R^r|e^{-\frac{i}{\hbar}H_kT}|\phi_K^k\rangle\;.
\end{gather}
As shown in the supplementary material, for $k\gg r$ and fixed $K$, $R$, this matrix element scales as
\begin{gather}
    \label{ukr}
    |U_{KR}^{kr}| \sim 2\sqrt{\frac{2}{\pi}} B_R^r \frac{|\mathcal{M}_2| f_{kr}}{k^3} \;, \\
     \label{fkr}
    f_{kr} =\frac{2r A_\lambda}{\sqrt{\left[1+\left(2k A_\lambda\right)^2\right]\left[1+\left(2r A_\lambda\right)^2\right]}}\;,
\end{gather}
where $|\mathcal{M}_2| = 2|J^{(2)}_{2(K-R)}\left(g\right)|$. For a fixed $A_\lambda$ and large enough $k$ and $r$ (such that $k A_\lambda, r A_\lambda\gg1$), the matrix elements of $U$ decay as
\begin{equation}
    \label{ukrkr}
    |U_{KR}^{kr}| \sim \frac{|\mathcal{M}_2|}{2\pi k^4 A_\lambda}\;.
\end{equation}
Therefore, for large $k$ and $r$, the matrix element $|U_{KR}^{kr}|$ decays super-exponentially fast with the center of mass $K$ momentum, due to the scaling of $\mathcal{M}_2$ which is controlled by the second order derivative of the Bessel function. The decay along the relative momentum $k$ direction however  is a power law $k^{-4}$, reflecting the presence of a  singular $\delta$-function interaction. This is our key result: super-exponential decay of the matrix element ensures the survival of dynamical localization for the center-of-mass momentum, while the power-law decay destroys it for the relative momentum. We expect that a smooth interaction function will instead lead to exponential matrix element decay in the relative momentum direction and survival of DL for weak enough interactions.

\begin{figure}
	\centering
	\includegraphics[width=1.0\columnwidth]{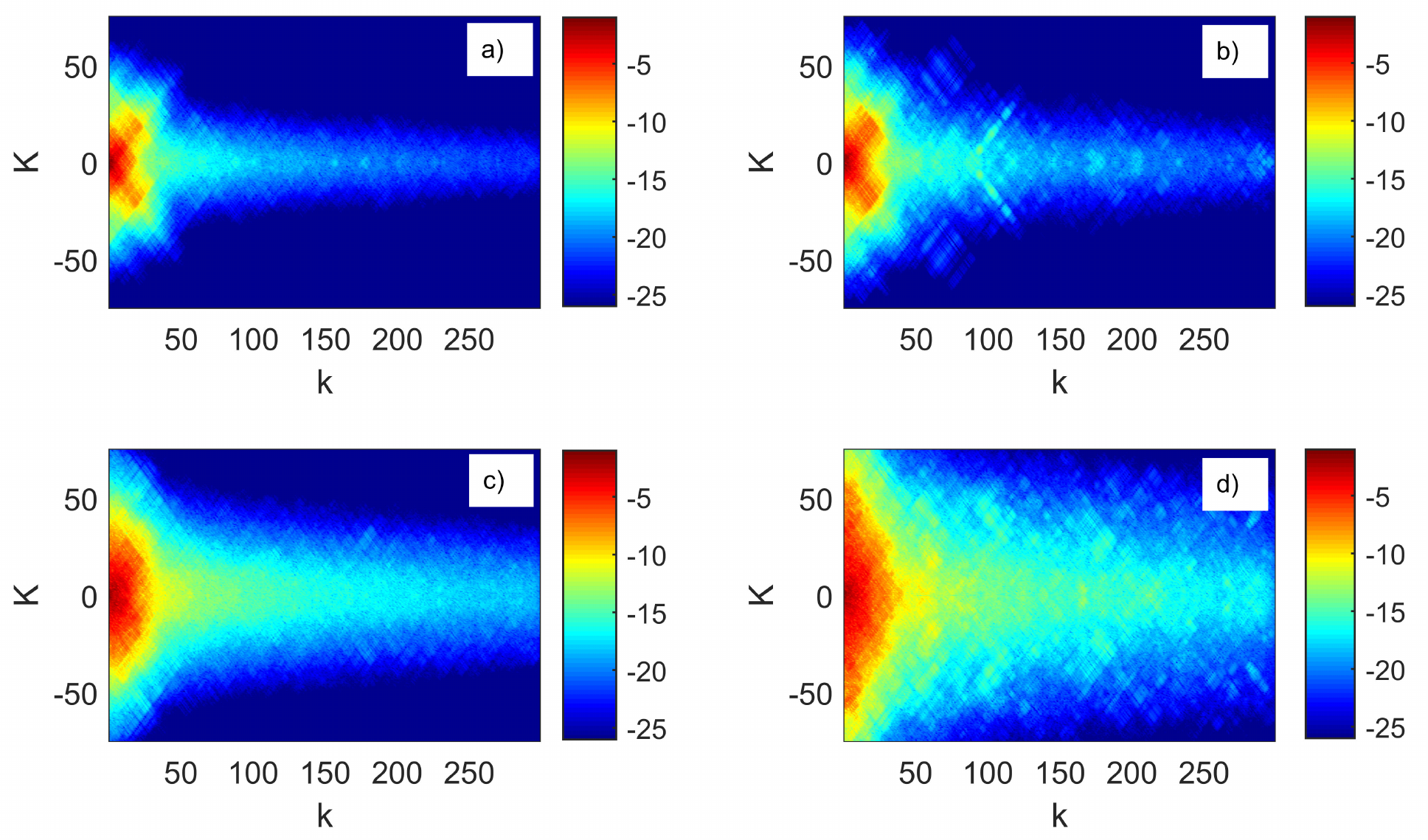}
	\caption{(Color online)  \textit{Top}: $|C_K^k(NT)|$ vs. $K$, $k$ with $A_\lambda=10$, $k_0=0.12$. \textit{Bottom}: $|C_K^k(NT)|$ vs. $K$, $k$, $A_\lambda=0.1$, $k_0=0.47$. a) and c) $N=100$, b) and d) $N=5000$.
Parameters: kicking strength $\xi=3$, driving period $T=1$ and momenta cutoffs $K_\text{max}=301$, $k_\text{max}=300$ and $K_0=0$.
}
	\label{fstateNine}
\end{figure}

We confirm the asymptotic decays with $K$ and $k$ and compute the matrix elements numerically using Eq.~\eqref{ukK} and the wave functions $|\phi_K^k\rangle$~\eqref{phikk}. The comparison of the numerical results to the asymptotic behavior is presented in Fig.~\ref{matrixdecay}. The top figure shows the decay of matrix element $U_{KR}^{kr}$ with relative momentum $k$ for several values of the coupling $A_\lambda$ indicated by colors. The other momenta $r, K, R$ are fixed. The power law fit $\log_{10} |U_{KR}^{kr}| = -4\log_{10} k - 0.2$ (the black line) agrees well with the numerical values of the matrix elements $U_{KR}^{kr}$ with a given $K,R,r$ and $k\gg r$. The small-$k$ dependence is not sensitive to the inverse interaction strength $A_\lambda$. The power-law decay of the matrix elements is not monotonic with $A_\lambda$: initially (blue, red and green curves) the prefactor is decreasing, however upon further decrease of $A_\lambda$ (magenta curve) it starts to increase and there is a non-monotonic intermediate part. This non-monotonicity can be explained from Eqs.~(\ref{ukr}-\ref{fkr}): for very small and very large $A_\lambda$, $f_{kr}$ is small, behaving as $2r A_\lambda$ and $1/(2k A_\lambda)$ respectively. Therefore, with decreasing $A_\lambda$, and for given $k,K,r,R$, the matrix element of $U$~\eqref{ukr} will first increase from a small value and then decrease back, which is precisely the non-monotonicity observed. For small $A_\lambda$ the power-laws have the same prefactors: this follows from~\eqref{fkr} - for a fixed $r$ and $r A_\lambda\ll1$, $f_{kr}\approx r/k$ is independent of $A_\lambda$. This is what we observe for the smallest values of $A_\lambda$ in Fig.~\ref{matrixdecay}.

The bottom plot in Fig.\ref{matrixdecay} shows the decay of the matrix elements as a function of $K$: a faster than exponential decay is observed, agreeing with the asymptotic behavior~\eqref{ukrkr}. The prefactor also shows non-monotonicity - the matrix elements initially increase with increasing $A_\lambda$ and later decrease - and has the same origin as above.

The impact of the decay properties of the matrix elements of $U$ is observed in the evolution of an intial state $|\psi(0)\rangle = |\phi_{K_0}^{k_0}\rangle$ with fixed momenta $K_0$ and $k_0$ according to Eq.~\eqref{timeevo}. This choice of intial state is well-suited to detect delocalisation in the momentum space. The final state after $N$ driving periods is $|\psi(NT)\rangle = \sum_{K,k} C_K^k(NT)|\phi_{K}^k\rangle$. We used the numerically evaluated $U$ to propagate the initial state in time.

Figure~\ref{fstateNine} shows the final state after $N=100$ - left column - and $N=5000$ - right column - driving periods for two different values $A_\lambda$, corresponding to moderately strong interaction between the rotors. The final state gets extended along the relative momentum $k$ direction. Also the extension more pronounced with decreasing $A_\lambda$.

\begin{figure}
	\centering
	\includegraphics[width=1.0\columnwidth]{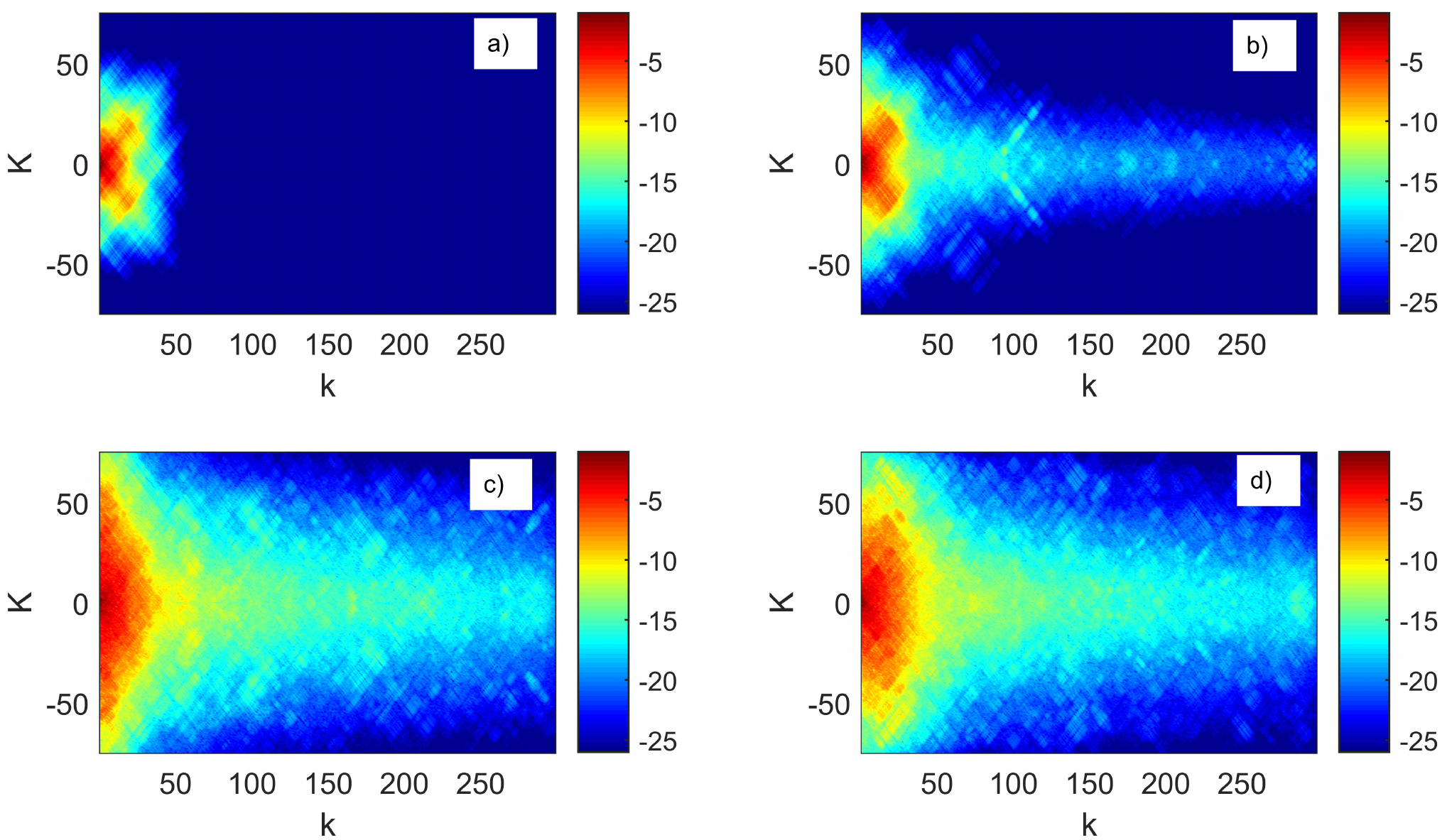}
	\caption{(Color online) $|C_K^k(NT)|$ vs. $K$, $k$ with momenta cutoffs $K_\text{max}=301$, $k_\text{max}=300$, $\xi=3$, $T=1$, $N=5000$. The interaction strengths for a) to d) $A_\lambda=10^{14}$, $A_\lambda=10$, $A_\lambda=0.1$, $A_\lambda=0.01$ respectively.}
	\label{fstatealmine}
\end{figure}

Figure~\ref{fstatealmine} shows the amplitude distribution of the final state after $N=5000$ driving periods for several values of $A_\lambda$. Fig.~\ref{fstatealmine}(a) shows the final state for the case of two essentially non-interacting rotors ($A_\lambda=10^{14}$). The final state is localized in both $K$ and $k$ momenta directions, displaying the dynamical localization of the non-interacting kicked rotor model. As the strength of the interaction is increasing, Figs.~\ref{fstatealmine}(b) and (c), the final state starts to extend along the $k$ direction. The interaction between the two rotors delocalizes the state  in the relative momentum $k$ direction. The localization length along the center of mass momentum $K$ direction also increases, as seen in Fig.~\ref{fstatealmine}(b) and (c). Fig.~\ref{fstatealmine}(d) shows the final state for strong interaction: compared with the (c) case, the extension of the final state along the $K$ direction has shrunk in a small momentum $k$ region. For very large momenta $k$ the amplitudes of the final state have values similar to the (c) case, since the matrix elements of $U$ become independent of $A_\lambda$ as we have discussed earlier.

\begin{figure}
	\centering
	\includegraphics[width=1.0\columnwidth]{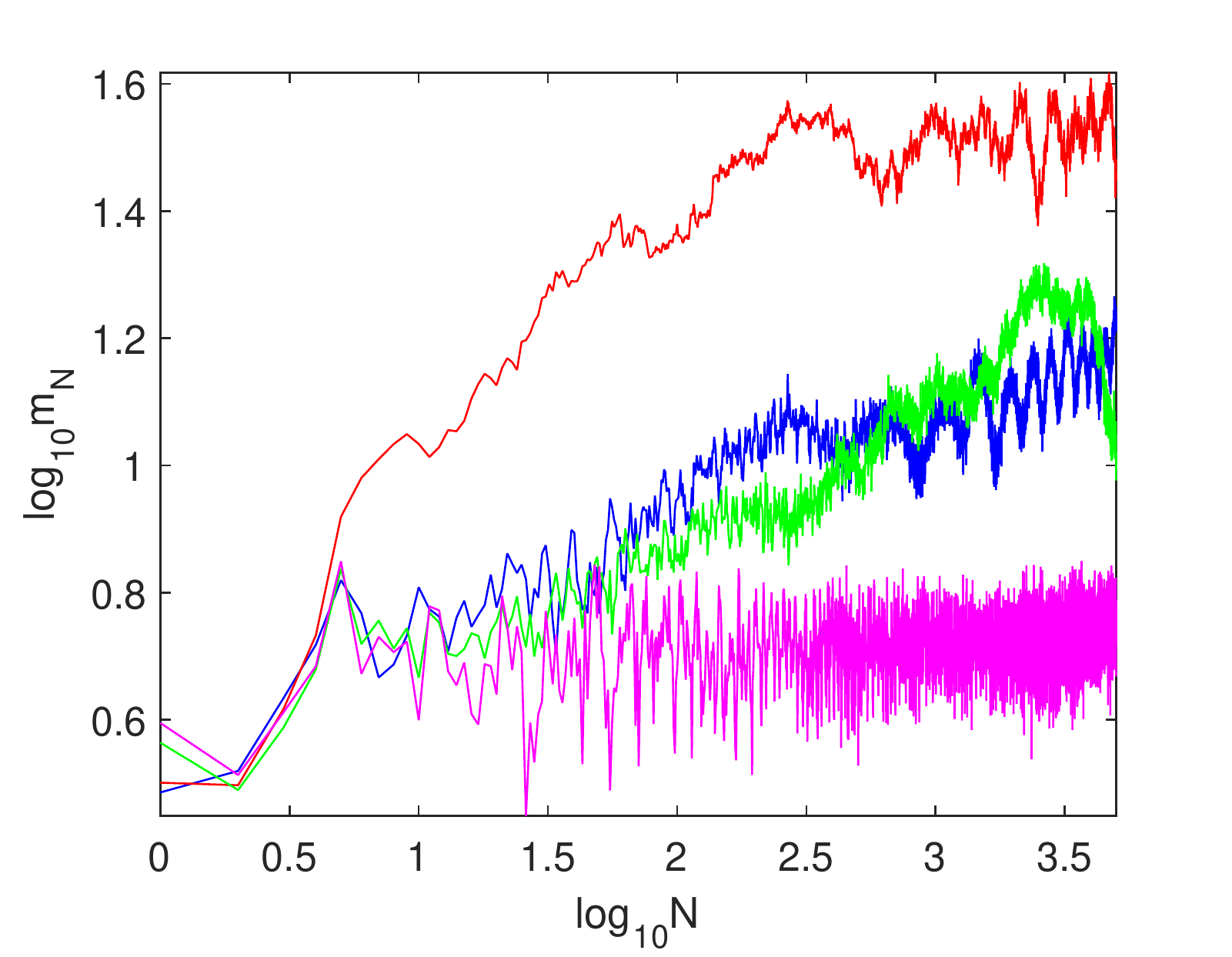}
	\caption{$\log_{10}m_N$ vs. $\log_{10}N$ with momenta cutoffs $K_\text{max}=301$, $k_\text{max}=300$, $\xi=3$, $T=1$ and several interaction strengths: blue line - $A_\lambda=0.01$, red line - $A_\lambda=0.1$, green line - $A_\lambda=1$, magenta line - $A_\lambda=10$.}
	\label{semine}
\end{figure}

In order to quantify the spreading of the initial state with $N$, we compute the evolution of the variance of the momenta:
\begin{gather}
    m_N = \sum_{K,k} \left[\left(K-\bar{K}\right)^2 + \left(k-\bar{k}\right)^2\right]\left|C_K^k(NT)\right|^2\\
    \bar{K} = \sum_{K,k} K\left|C_K^k(NT)\right|^2\\
    \bar{k} = \sum_{K,k} k\left|C_K^k(NT)\right|^2.
\end{gather}
Figure~\ref{semine} shows the evolution of the variance with $N$ for several strengths of the interaction $A_\lambda$. For all the values of the interaction except $A_\lambda=10$ the variance has a clearly increasing trend spanning several orders of magnitude in $N$, therefore signalizing delocalization along the relative momentum $k$ direction. The non-monotonic dependence of the variance on $A_\lambda$ has the same origin as the non-monotonic dependence of the matrix elements of $U$ that we discussed above.

In conclusion, we studied dynamical properties of two interacting kicked quantum rotors. Due to the non-analyticity of the $\delta$-function interaction, the matrix elements of the time evolution operator exhibit different decay behaviors in center of mass and relative momentum directions. Along the center of mass momentum direction, matrix elements decay super-exponentially. Along the relative momentum direction, matrix elements decay as a power-law with the exponent $4$. As a result the center of mass motion remains localized like in the non-interacting case, while the relative motion becomes extended. This effect should lead to a destruction of DL in interacting ultracold atomic gases and be easily observable in a setup similar to the one used in~\cite{moore1994observation} using Feshbach resonances. To analyze this, we need to consider many interacting atoms and study the highly complex case of many body interactions for quantum kicked rotors. While this is still a challenging task, we refer to mean field treatments of this case in \cite{shepelyansky93delocalization,gligoric2011interactions} which demonstrate the complete destruction of dynamical localization as well.

\bibliographystyle{apsrev4-1}
\bibliography{driven}

\begin{thebibliography}{18}%
\makeatletter
\providecommand \@ifxundefined [1]{%
 \@ifx{#1\undefined}
}%
\providecommand \@ifnum [1]{%
 \ifnum #1\expandafter \@firstoftwo
 \else \expandafter \@secondoftwo
 \fi
}%
\providecommand \@ifx [1]{%
 \ifx #1\expandafter \@firstoftwo
 \else \expandafter \@secondoftwo
 \fi
}%
\providecommand \natexlab [1]{#1}%
\providecommand \enquote  [1]{``#1''}%
\providecommand \bibnamefont  [1]{#1}%
\providecommand \bibfnamefont [1]{#1}%
\providecommand \citenamefont [1]{#1}%
\providecommand \href@noop [0]{\@secondoftwo}%
\providecommand \href [0]{\begingroup \@sanitize@url \@href}%
\providecommand \@href[1]{\@@startlink{#1}\@@href}%
\providecommand \@@href[1]{\endgroup#1\@@endlink}%
\providecommand \@sanitize@url [0]{\catcode `\\12\catcode `\$12\catcode
  `\&12\catcode `\#12\catcode `\^12\catcode `\_12\catcode `\%12\relax}%
\providecommand \@@startlink[1]{}%
\providecommand \@@endlink[0]{}%
\providecommand \url  [0]{\begingroup\@sanitize@url \@url }%
\providecommand \@url [1]{\endgroup\@href {#1}{\urlprefix }}%
\providecommand \urlprefix  [0]{URL }%
\providecommand \Eprint [0]{\href }%
\providecommand \doibase [0]{http://dx.doi.org/}%
\providecommand \selectlanguage [0]{\@gobble}%
\providecommand \bibinfo  [0]{\@secondoftwo}%
\providecommand \bibfield  [0]{\@secondoftwo}%
\providecommand \translation [1]{[#1]}%
\providecommand \BibitemOpen [0]{}%
\providecommand \bibitemStop [0]{}%
\providecommand \bibitemNoStop [0]{.\EOS\space}%
\providecommand \EOS [0]{\spacefactor3000\relax}%
\providecommand \BibitemShut  [1]{\csname bibitem#1\endcsname}%
\let\auto@bib@innerbib\@empty
\bibitem [{\citenamefont {Chirikov}(1979)}]{chirikov1979universal}%
  \BibitemOpen
  \bibfield  {author} {\bibinfo {author} {\bibfnamefont {B.~V.}\ \bibnamefont
  {Chirikov}},\ }\href {\doibase
  http://dx.doi.org/10.1016/0370-1573(79)90023-1} {\bibfield  {journal}
  {\bibinfo  {journal} {Phys. Rep.}\ }\textbf {\bibinfo {volume} {52}},\
  \bibinfo {pages} {263 } (\bibinfo {year} {1979})}\BibitemShut {NoStop}%
\bibitem [{\citenamefont {Izrailev}(1990)}]{izrailev1990simple}%
  \BibitemOpen
  \bibfield  {author} {\bibinfo {author} {\bibfnamefont {F.~M.}\ \bibnamefont
  {Izrailev}},\ }\href {\doibase
  http://dx.doi.org/10.1016/0370-1573(90)90067-C} {\bibfield  {journal}
  {\bibinfo  {journal} {Phys. Rep.}\ }\textbf {\bibinfo {volume} {196}},\
  \bibinfo {pages} {299 } (\bibinfo {year} {1990})}\BibitemShut {NoStop}%
\bibitem [{\citenamefont {Casati}\ \emph {et~al.}(1979)\citenamefont {Casati},
  \citenamefont {Chirikov}, \citenamefont {Izraelev},\ and\ \citenamefont
  {Ford}}]{casati1979stochastic}%
  \BibitemOpen
  \bibfield  {author} {\bibinfo {author} {\bibfnamefont {G.}~\bibnamefont
  {Casati}}, \bibinfo {author} {\bibfnamefont {B.~V.}\ \bibnamefont
  {Chirikov}}, \bibinfo {author} {\bibfnamefont {F.~M.}\ \bibnamefont
  {Izraelev}}, \ and\ \bibinfo {author} {\bibfnamefont {J.}~\bibnamefont
  {Ford}},\ }\enquote {\bibinfo {title} {Stochastic behavior in classical and
  quantum hamiltonian systems: Volta memorial conference, como, 1977},}\ \
  (\bibinfo  {publisher} {Springer Berlin Heidelberg},\ \bibinfo {address}
  {Berlin, Heidelberg},\ \bibinfo {year} {1979})\BibitemShut {NoStop}%
\bibitem [{\citenamefont {Galvez}\ \emph {et~al.}(1988)\citenamefont {Galvez},
  \citenamefont {Sauer}, \citenamefont {Moorman}, \citenamefont {Koch},\ and\
  \citenamefont {Richards}}]{galvez1988microwave}%
  \BibitemOpen
  \bibfield  {author} {\bibinfo {author} {\bibfnamefont {E.~J.}\ \bibnamefont
  {Galvez}}, \bibinfo {author} {\bibfnamefont {B.~E.}\ \bibnamefont {Sauer}},
  \bibinfo {author} {\bibfnamefont {L.}~\bibnamefont {Moorman}}, \bibinfo
  {author} {\bibfnamefont {P.~M.}\ \bibnamefont {Koch}}, \ and\ \bibinfo
  {author} {\bibfnamefont {D.}~\bibnamefont {Richards}},\ }\href {\doibase
  10.1103/PhysRevLett.61.2011} {\bibfield  {journal} {\bibinfo  {journal}
  {Phys. Rev. Lett.}\ }\textbf {\bibinfo {volume} {61}},\ \bibinfo {pages}
  {2011} (\bibinfo {year} {1988})}\BibitemShut {NoStop}%
\bibitem [{\citenamefont {Bayfield}\ \emph {et~al.}(1989)\citenamefont
  {Bayfield}, \citenamefont {Casati}, \citenamefont {Guarneri},\ and\
  \citenamefont {Sokol}}]{bayfield1989localization}%
  \BibitemOpen
  \bibfield  {author} {\bibinfo {author} {\bibfnamefont {J.~E.}\ \bibnamefont
  {Bayfield}}, \bibinfo {author} {\bibfnamefont {G.}~\bibnamefont {Casati}},
  \bibinfo {author} {\bibfnamefont {I.}~\bibnamefont {Guarneri}}, \ and\
  \bibinfo {author} {\bibfnamefont {D.~W.}\ \bibnamefont {Sokol}},\ }\href
  {\doibase 10.1103/PhysRevLett.63.364} {\bibfield  {journal} {\bibinfo
  {journal} {Phys. Rev. Lett.}\ }\textbf {\bibinfo {volume} {63}},\ \bibinfo
  {pages} {364} (\bibinfo {year} {1989})}\BibitemShut {NoStop}%
\bibitem [{\citenamefont {Moore}\ \emph {et~al.}(1994)\citenamefont {Moore},
  \citenamefont {Robinson}, \citenamefont {Bharucha}, \citenamefont
  {Williams},\ and\ \citenamefont {Raizen}}]{moore1994observation}%
  \BibitemOpen
  \bibfield  {author} {\bibinfo {author} {\bibfnamefont {F.~L.}\ \bibnamefont
  {Moore}}, \bibinfo {author} {\bibfnamefont {J.~C.}\ \bibnamefont {Robinson}},
  \bibinfo {author} {\bibfnamefont {C.}~\bibnamefont {Bharucha}}, \bibinfo
  {author} {\bibfnamefont {P.~E.}\ \bibnamefont {Williams}}, \ and\ \bibinfo
  {author} {\bibfnamefont {M.~G.}\ \bibnamefont {Raizen}},\ }\href {\doibase
  10.1103/PhysRevLett.73.2974} {\bibfield  {journal} {\bibinfo  {journal}
  {Phys. Rev. Lett.}\ }\textbf {\bibinfo {volume} {73}},\ \bibinfo {pages}
  {2974} (\bibinfo {year} {1994})}\BibitemShut {NoStop}%
\bibitem [{\citenamefont {Bitter}\ and\ \citenamefont
  {Milner}(2016)}]{bitter2016experimental}%
  \BibitemOpen
  \bibfield  {author} {\bibinfo {author} {\bibfnamefont {M.}~\bibnamefont
  {Bitter}}\ and\ \bibinfo {author} {\bibfnamefont {V.}~\bibnamefont
  {Milner}},\ }\href@noop {} {\  (\bibinfo {year} {2016})},\ \Eprint
  {http://arxiv.org/abs/1603.06918} {arXiv:1603.06918 [quant-ph]} \BibitemShut
  {NoStop}%
\bibitem [{\citenamefont {Fishman}\ \emph {et~al.}(1982)\citenamefont
  {Fishman}, \citenamefont {Grempel},\ and\ \citenamefont
  {Prange}}]{fishman1982chaos}%
  \BibitemOpen
  \bibfield  {author} {\bibinfo {author} {\bibfnamefont {S.}~\bibnamefont
  {Fishman}}, \bibinfo {author} {\bibfnamefont {D.~R.}\ \bibnamefont
  {Grempel}}, \ and\ \bibinfo {author} {\bibfnamefont {R.~E.}\ \bibnamefont
  {Prange}},\ }\href {\doibase 10.1103/PhysRevLett.49.509} {\bibfield
  {journal} {\bibinfo  {journal} {Phys. Rev. Lett.}\ }\textbf {\bibinfo
  {volume} {49}},\ \bibinfo {pages} {509} (\bibinfo {year} {1982})}\BibitemShut
  {NoStop}%
\bibitem [{\citenamefont {Keser}\ \emph {et~al.}(2015)\citenamefont {Keser},
  \citenamefont {Ganeshan}, \citenamefont {Refael},\ and\ \citenamefont
  {Galitski}}]{keser2015dynamical}%
  \BibitemOpen
  \bibfield  {author} {\bibinfo {author} {\bibfnamefont {A.~C.}\ \bibnamefont
  {Keser}}, \bibinfo {author} {\bibfnamefont {S.}~\bibnamefont {Ganeshan}},
  \bibinfo {author} {\bibfnamefont {G.}~\bibnamefont {Refael}}, \ and\ \bibinfo
  {author} {\bibfnamefont {V.}~\bibnamefont {Galitski}},\ }\href@noop {} {\
  (\bibinfo {year} {2015})},\ \Eprint {http://arxiv.org/abs/1506.05455}
  {arxiv:1506.05455 [cond-mat.dis-nn]} \BibitemShut {NoStop}%
\bibitem [{\citenamefont {Rozenbaum}\ and\ \citenamefont
  {Galitski}(2016)}]{rozenbaum2016dynamical}%
  \BibitemOpen
  \bibfield  {author} {\bibinfo {author} {\bibfnamefont {E.~B.}\ \bibnamefont
  {Rozenbaum}}\ and\ \bibinfo {author} {\bibfnamefont {V.}~\bibnamefont
  {Galitski}},\ }\href@noop {} {\  (\bibinfo {year} {2016})},\ \Eprint
  {http://arxiv.org/abs/1602.04425} {arXiv:1602.04425 [cond-mat.dis-nn]}
  \BibitemShut {NoStop}%
\bibitem [{\citenamefont {Nag}\ \emph {et~al.}(2005)\citenamefont {Nag},
  \citenamefont {Ghosh},\ and\ \citenamefont {Lahiri}}]{nag2005quantum}%
  \BibitemOpen
  \bibfield  {author} {\bibinfo {author} {\bibfnamefont {S.}~\bibnamefont
  {Nag}}, \bibinfo {author} {\bibfnamefont {G.}~\bibnamefont {Ghosh}}, \ and\
  \bibinfo {author} {\bibfnamefont {A.}~\bibnamefont {Lahiri}},\ }\href
  {\doibase http://dx.doi.org/10.1016/j.physd.2005.04.008} {\bibfield
  {journal} {\bibinfo  {journal} {Physica D}\ }\textbf {\bibinfo {volume}
  {204}},\ \bibinfo {pages} {110 } (\bibinfo {year} {2005})}\BibitemShut
  {NoStop}%
\bibitem [{\citenamefont {Adachi}\ \emph {et~al.}(1988)\citenamefont {Adachi},
  \citenamefont {Toda},\ and\ \citenamefont {Ikeda}}]{adachi1988quantum}%
  \BibitemOpen
  \bibfield  {author} {\bibinfo {author} {\bibfnamefont {S.}~\bibnamefont
  {Adachi}}, \bibinfo {author} {\bibfnamefont {M.}~\bibnamefont {Toda}}, \ and\
  \bibinfo {author} {\bibfnamefont {K.}~\bibnamefont {Ikeda}},\ }\href
  {\doibase 10.1103/PhysRevLett.61.659} {\bibfield  {journal} {\bibinfo
  {journal} {Phys. Rev. Lett.}\ }\textbf {\bibinfo {volume} {61}},\ \bibinfo
  {pages} {659} (\bibinfo {year} {1988})}\BibitemShut {NoStop}%
\bibitem [{\citenamefont {Toloui}\ and\ \citenamefont
  {Ballentine}(2009)}]{toloui2009quantum}%
  \BibitemOpen
  \bibfield  {author} {\bibinfo {author} {\bibfnamefont {B.}~\bibnamefont
  {Toloui}}\ and\ \bibinfo {author} {\bibfnamefont {L.~E.}\ \bibnamefont
  {Ballentine}},\ }\href@noop {} {\  (\bibinfo {year} {2009})},\ \Eprint
  {http://arxiv.org/abs/0903.4632} {arXiv:0903.4632} \BibitemShut {NoStop}%
\bibitem [{\citenamefont {Cheng}\ \emph {et~al.}(2010)\citenamefont {Cheng},
  \citenamefont {Rudolf}, \citenamefont {Paul},\ and\ \citenamefont
  {Eite}}]{chin2010feshbach}%
  \BibitemOpen
  \bibfield  {author} {\bibinfo {author} {\bibfnamefont {C.}~\bibnamefont
  {Cheng}}, \bibinfo {author} {\bibfnamefont {G.}~\bibnamefont {Rudolf}},
  \bibinfo {author} {\bibfnamefont {J.}~\bibnamefont {Paul}}, \ and\ \bibinfo
  {author} {\bibfnamefont {T.}~\bibnamefont {Eite}},\ }\href {\doibase
  http://dx.doi.org/10.1103/RevModPhys.82.1225} {\bibfield  {journal} {\bibinfo
   {journal} {Rev. Mod. Phys.}\ }\textbf {\bibinfo {volume} {823}},\ \bibinfo
  {pages} {1225} (\bibinfo {year} {2010})}\BibitemShut {NoStop}%
\bibitem [{\citenamefont {Lieb}\ and\ \citenamefont
  {Liniger}(1963)}]{lieb1963exact}%
  \BibitemOpen
  \bibfield  {author} {\bibinfo {author} {\bibfnamefont {E.~H.}\ \bibnamefont
  {Lieb}}\ and\ \bibinfo {author} {\bibfnamefont {W.}~\bibnamefont {Liniger}},\
  }\href {\doibase 10.1103/PhysRev.130.1605} {\bibfield  {journal} {\bibinfo
  {journal} {Phys. Rev.}\ }\textbf {\bibinfo {volume} {130}},\ \bibinfo {pages}
  {1605} (\bibinfo {year} {1963})}\BibitemShut {NoStop}%
\bibitem [{\citenamefont {Grifoni}\ and\ \citenamefont
  {H{\"a}nggi}(1998)}]{grifoni1998driven}%
  \BibitemOpen
  \bibfield  {author} {\bibinfo {author} {\bibfnamefont {M.}~\bibnamefont
  {Grifoni}}\ and\ \bibinfo {author} {\bibfnamefont {P.}~\bibnamefont
  {H{\"a}nggi}},\ }\href {\doibase
  http://dx.doi.org/10.1016/S0370-1573(98)00022-2} {\bibfield  {journal}
  {\bibinfo  {journal} {Phys. Rep.}\ }\textbf {\bibinfo {volume} {304}},\
  \bibinfo {pages} {229 } (\bibinfo {year} {1998})}\BibitemShut {NoStop}%
\bibitem [{\citenamefont {Shepelyansky}(1993)}]{shepelyansky93delocalization}%
  \BibitemOpen
  \bibfield  {author} {\bibinfo {author} {\bibfnamefont {D.~L.}\ \bibnamefont
  {Shepelyansky}},\ }\href {\doibase 10.1103/PhysRevLett.70.1787} {\bibfield
  {journal} {\bibinfo  {journal} {Phys. Rev. Lett.}\ }\textbf {\bibinfo
  {volume} {70}},\ \bibinfo {pages} {1787} (\bibinfo {year}
  {1993})}\BibitemShut {NoStop}%
\bibitem [{\citenamefont {Gligorić}\ \emph {et~al.}(2011)\citenamefont
  {Gligorić}, \citenamefont {Bodyfelt},\ and\ \citenamefont
  {Flach}}]{gligoric2011interactions}%
  \BibitemOpen
  \bibfield  {author} {\bibinfo {author} {\bibfnamefont {G.}~\bibnamefont
  {Gligorić}}, \bibinfo {author} {\bibfnamefont {J.~D.}\ \bibnamefont
  {Bodyfelt}}, \ and\ \bibinfo {author} {\bibfnamefont {S.}~\bibnamefont
  {Flach}},\ }\href {http://stacks.iop.org/0295-5075/96/i=3/a=30004} {\bibfield
   {journal} {\bibinfo  {journal} {EPL}\ }\textbf {\bibinfo {volume} {96}},\
  \bibinfo {pages} {30004} (\bibinfo {year} {2011})}\BibitemShut {NoStop}%
\end{thebibliography}%

\end{document}